\documentclass[12pt]{article}

\usepackage{graphicx}
\usepackage{natbib}
\usepackage[table,usenames,dvipsnames]{xcolor}
\usepackage[margin=1.5cm]{geometry}
\usepackage[hypertexnames=false,colorlinks=true,breaklinks]{hyperref}
\usepackage{rotating}
\usepackage{verbatim}
\usepackage{lineno}
\usepackage{caption}
\usepackage{url}
\usepackage[toc,nonumberlist,numberedsection=autolabel]{glossaries}
\usepackage{amsmath}

\newglossaryentry{PAC} {
    name={PAC},
    description={phase-amplitude coupling. Coupling between the phase of
low-frequency oscillations and the amplitude of high-frequency ones}
}

\newglossaryentry{ERP} {
    name={ERP},
    description={event-related potential. Mean of voltages recorded at a single
electrode and aligned with respect to an event of interest (e.g., the
initiation of the production of \glspl{CVS}}
}

\newglossaryentry{MI} {
    name={MI},
    description={modulation index quantifying the strength of phase-amplitude coupling}
}

\newglossaryentry{TW} {
    name={TW},
    description={traveling wave}
}

\newglossaryentry{CVS} {
    name={CVS},
    description={consonant vowel syllable}
}

\newglossaryentry{WE} {
    name={WE},
    description={wave event. A sample time with a significant linear relation
    between phase differences of electrodes with respect to the reference
    electrode and distances of these electrodes to the reference electrode}
}

\newglossaryentry{CWE} {
    name={CWE},
    description={contiguous wave event.  sequence of un-interrupted of \glspl{WE} such that \glspl{WE} are detected in all samples times from the sample time of the first \gls{WE} to the sample time of the last \gls{WE}}
}

\newglossaryentry{vSMC} {
    name={vSMC},
    description={ventral sensorimotor cortex. A brain region that controls the
vocal articulators}
}

\newglossaryentry{ECoG} {
    name={ECoG},
    description={electrocorticography. A brain recording modality that measures
potentials directly from the cortical surface}
}

\newglossaryentry{MEG} {
    name={MEG},
    description={magnetoencephalography. A brain recording modality that 
measures magnetic fields from the scalp}
}

\newglossaryentry{EEG} {
    name={EEG},
    description={electroencephalography. A brain recording modality that 
measures electrical potentials from the scalp}
}

\newglossaryentry{PLI} {
    name={PLI},
    description={phase-locking index. A measure of alignment among multiple phases}
}

\newglossaryentry{ERSP} {
    name={ERSP},
    description={event-related spectral perturbation. An average measure of evoked power across trials}
}

\newglossaryentry{cvsPhase} {
    name={CVS phase},
    description={phase of filtered-voltage oscillation at which a \gls{CVS} is
initiated}
}

\makeglossaries

\begin{document}


\title{Traveling waves appear and disappear in unison with produced speech}

\author{Joaqu\'{i}n Rapela\thanks{rapela@ucsd.edu}}
\maketitle

\abstract{

In \citet{rapelaInPrepTWsInSpeech} we reported traveling waves (\glspl{TW}) on
electrocorticographic (\gls{ECoG}) recordings from an epileptic subject over
speech processing brain regions, while the subject rhythmically produced
consonant-vowel syllables (\glspl{CVS}). In \citet{rapelaInPrepSyncTWs} we
showed that \glspl{TW} are precisely synchronized, in dynamical systems terms,
to these productions.
Here we show that \glspl{TW} do not occur continuously, but tend to appear
before the initiation of \glspl{CVS} and tend to disappear before their
termination.  During moments of silence, between productions of \glspl{CVS},
\glspl{TW} tend to reverse direction.
To our knowledge, this is the first study showing \glspl{TW} related to the
production of speech and, more generally, the first report of behavioral
correlates of mesoscale \glspl{TW} in behaving humans.

}

\pagebreak


\pagebreak
\setcounter{page}{1}

\section{Introduction}

Findings of spatio-temporally organized neural activity (i.e., traveling
waves, \glspl{TW}) are not new. However, their functional role is still
unclear.

In non-human animals \glspl{TW} have been related to the perception of visual
stimuli by monkeys \citep[e.g.,][]{benucciEtAl07} and by turtles
\citep[e.g.,][]{prechtlEtAl97}, and to movement preparation by monkeys
\citep[e.g.,][]{rubinoEtAl06}.
In humans, macroscopic \glspl{TW} have been associated to reaction speeds
\citep[e.g.,][]{pattenEtAl12} and to memory  consolidation during sleep
\cite[e.g.,][]{mullerEtAl16}, among others.
However, to our knowledge, no study has reported \glspl{TW} during the
production of speech or has found behavioral correlates of mesoscopic
\glspl{TW} in behaving humans.

In \citet{rapelaInPrepTWsInSpeech} we reported the existence of \glspl{TW} in
a speech processing brain region. In \citet{rapelaInPrepSyncTWs} we showed
that, at times of \gls{CVS} transitions, \glspl{TW} are precisely
synchronized, in dynamical systems terms, to the production of \glspl{CVS}.
Here we show that \glspl{TW} tend to appear before initiations of productions
of \glspl{CVS}, tend to disappear before the their terminations, and sometimes
reverse direction during moments of silence.

\section{Results}

\begin{figure}
\begin{center}
\includegraphics[width=6in]{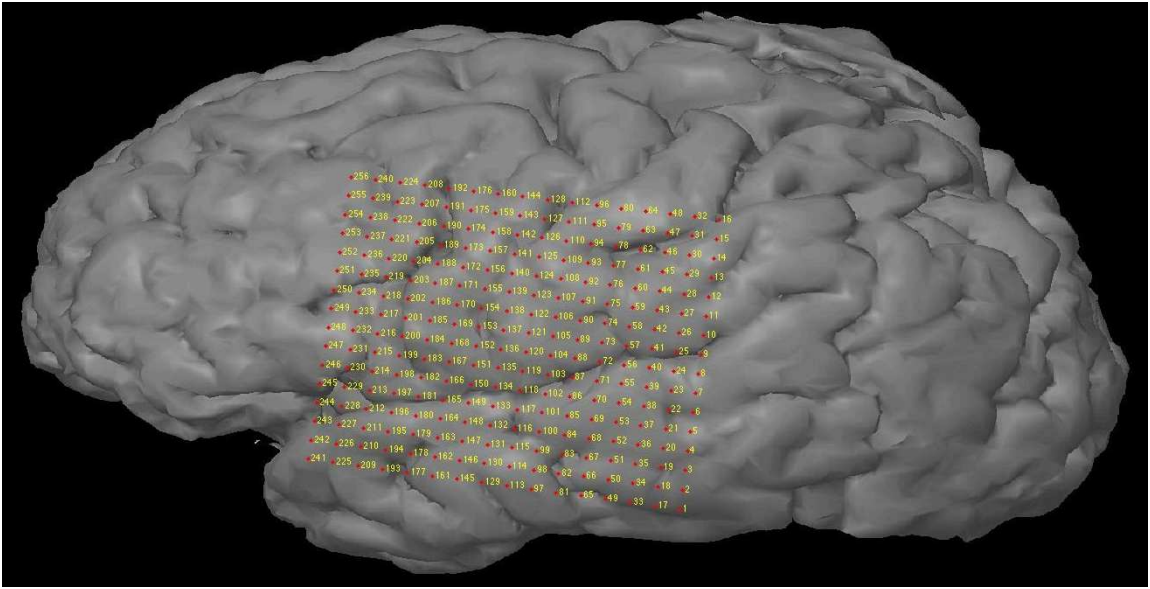}
\end{center}

\caption{Grid of ECoG electrodes superimposed on an MRI reconstruction of a
subjects' brain.}

\label{fig:grid}
\end{figure}

We describe findings from the analysis of \glspl{TW} in recording session
EC2\_B105 of subject EC2. We report results from the analysis of \glspl{TW}
moving over the ventral sensorimotor cortex between electrodes 135 and 142
(Figure~\ref{fig:grid}) at the median frequency of \gls{CVS} production of
this subject, 0.62~Hz.

We say that there is a wave event (\gls{WE}) at a sample time when recorded
neural activity at this time is spatiotemporally organized as a \gls{TW}
(Section~\ref{sec:we}).
We say that there is a contiguous wave event (\gls{CWE}) from time $t_0$ to
time $t_1$ when \glspl{WE} are detected at all sample times between $t_0$ and
$t_1$ (Section~\ref{sec:cwe}).

As we quantify below, \glspl{TW} moving in the ventro-dorsal direction (from
electrode 135 to 142) were not continuous, but tended to appear only during
the production of \glspl{CVS}.  The start and end times of \glspl{CWE} was
aligned to initiations and terminations of \glspl{CVS} with a sub-second
precision.  Also, we observed few \glspl{TW} moving in the dorso-ventral
direction (from electrode 142 to 135), which appeared most often during
periods of silence between productions of \glspl{CVS}.

\begin{figure} \begin{center}
\href{https://sccn.ucsd.edu/~rapela/eddyECOGSpeech/analysis/syncTWsAndCVSs/figures/EC2_B105/contiguousWaveEventsWithoutPhaseUnwrappingSignificance0.01rTheshold0.85FilteredFrom0.40To0.80Order02ZScored0SaveFromTime0.00ToTime700.00DT0.10FromElec135ToElec142.html}{\includegraphics[width=6in]{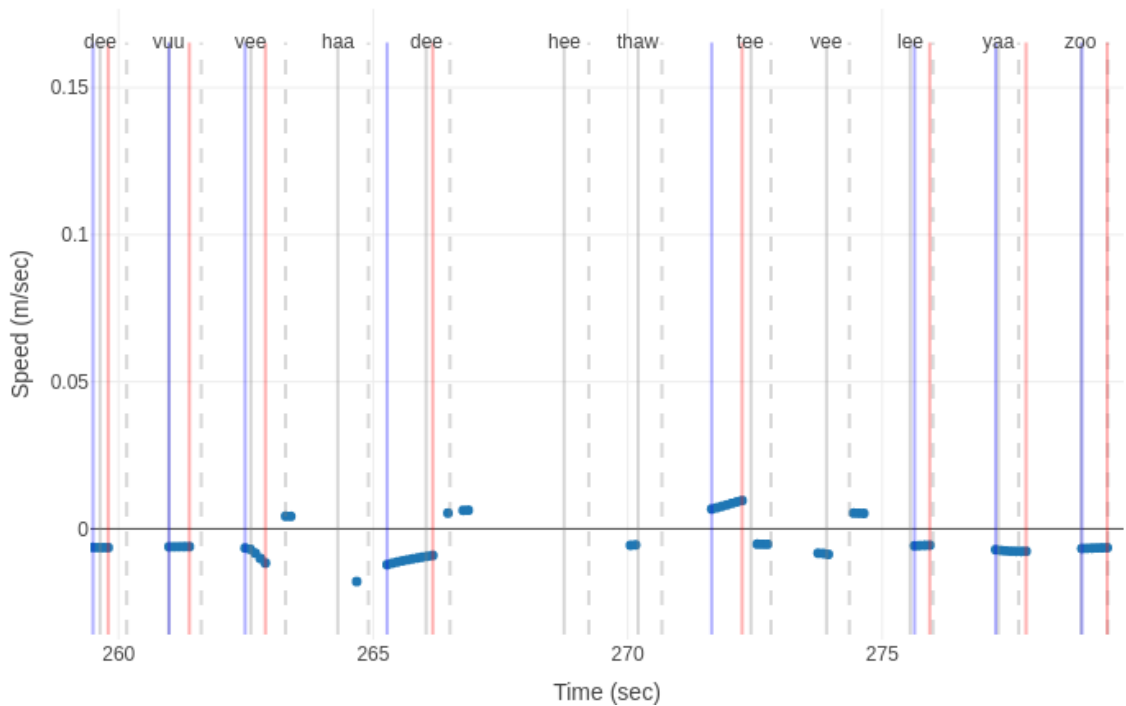}}
\caption{
During the production of \glspl{CVS} traveling waves tend to propagate in the
ventro-dorsal direction (i.e., have negative speed). They tend to appear
before the initiation of \glspl{CVS} and disappear before their termination.
In moments of silence, \glspl{TW} tend to reverse direction and propagate in
the dorso-ventral direction (i.e., have positive speed).
Gray vertical lines indicate behavior; solid and dashed lines mark initiations
and terminations of productions of \glspl{CVS}, respectively.
Colored dots and vertical lines indicate physiology; blue dots mark
\glspl{WE}, blue and red vertical lines indicate starts and ends,
respectively, of \glspl{CWE}.
In most cases \glspl{CWE} start and end before the initiation and termination
of \glspl{CVS} (i.e., blue vertical lines appear a little before solid gray
                      vertical lines and red vertical lines appear a little
                      before gray dashed lines).
Click on the figure to access its online version plotting the full ten-minutes
recording session.
} 

\label{fig:significantSpeeds}
\end{center}
\end{figure}

The abscissa of a blue dot in Figure~\ref{fig:significantSpeeds} marks the
time at which a \gls{WE} was detected, and the ordinate gives the speed of the
corresponding \gls{TW}.
Blue/red lines in Figure~\ref{fig:significantSpeeds} mark start/end times of
\glspl{CWE}.
Solid and dashed gray lines in Figure~\ref{fig:significantSpeeds} indicate
initiations and terminations, respectively, of the \glspl{CVS} at the top of
the plot.
For instance, the blue and red lines at 326.3 and 327.1~seconds mark the start
and end times of a \gls{CWE} and the solid and dotted black lines at 326.4 and
327.2~seconds mark the initiation and terminations of the \gls{CVS}
\texttt{zaa}.
We see that for ventro-dorsal \glspl{TW} (i.e., \glspl{CWE} with negative
speed) tend to start a little before initiations of \glspl{CWE} (i.e., blue
vertical lines corresponding to blue dots with negative speed appear a little
to the left of solid vertical gray lines) and end a little before terminations
of \glspl{CWE} (i.e., red vertical lines corresponding to blue dots with
negative speed appear a little to the left of dashed vertial gray lines).

We observed 1293 
ventro-dorsal \glspl{WE} (i.e., moving from
electrode 135 to 142). The proportion of significant such events occurring
between 0.3~seconds before the start time of the production of a \gls{CVS} and
the end time of this production was 86\%, and this proportion was
significantly different from chance (p\textless 1e-4, permutation test,
Section~\ref{sec:permutationTests}).
We observed 116 dorso-ventral \glspl{WE} and these events 
occurred most often during periods of silence.  The proportion of significant
dorso-ventral \glspl{WE}
in periods of silence was 90\%, and this proportion was significantly
different from chance (p=0.006, permutation test,
Section~\ref{sec:permutationTests}).
That is, ventro-dorsal \glspl{TW} tended to occur during \glspl{CVS}
productions and these \glspl{TW} tended to reverse direction during silence.

\begin{figure}
\centering
\includegraphics[width=4in]{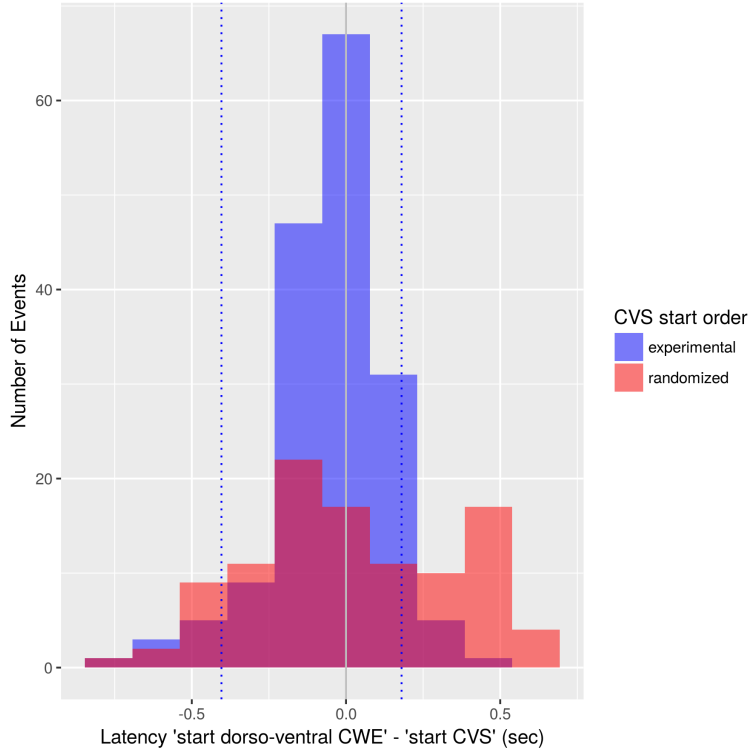}

\caption{
Histograms of latencies between the start of ventro-dorsal \glspl{CWE} and the
initiation of \glspl{CVS}. The blue and red histograms correspond to
experimental and surrogate datasets (Section~\ref{sec:surrogateDatasets}). 
Dotted blue vertical lines indicate 5\% and 95\% percentiles of the
experimental dataset histogram. 90\% of \glspl{CWE} started between 0.4~sec
before and 0.18~sec after the initiation of a \gls{CVS}.
The mean latency in the experimental dataset was -0.06~sec, which was
significantly smaller than chance (p=0.0185, permutation test,
Section~\ref{sec:permutationTests}), suggesting that on average \glspl{CWE} started
before the initiation of productions of \glspl{CVS}.
The variance of latencies in the experimental dataset was 0.03~sec$^2$, which
was significantly smaller than chance (p\textless 1e-4, permutation test,
Section~\ref{sec:permutationTests}), suggesting that on average \glspl{CWE} started
closer to the initiation of productions of \glspl{CVS} than expected by
chance.
}

\label{fig:histLatenciesInitiationPosSpeedCWEs}
\end{figure}

\begin{figure}
\begin{center}
\includegraphics[width=4in]{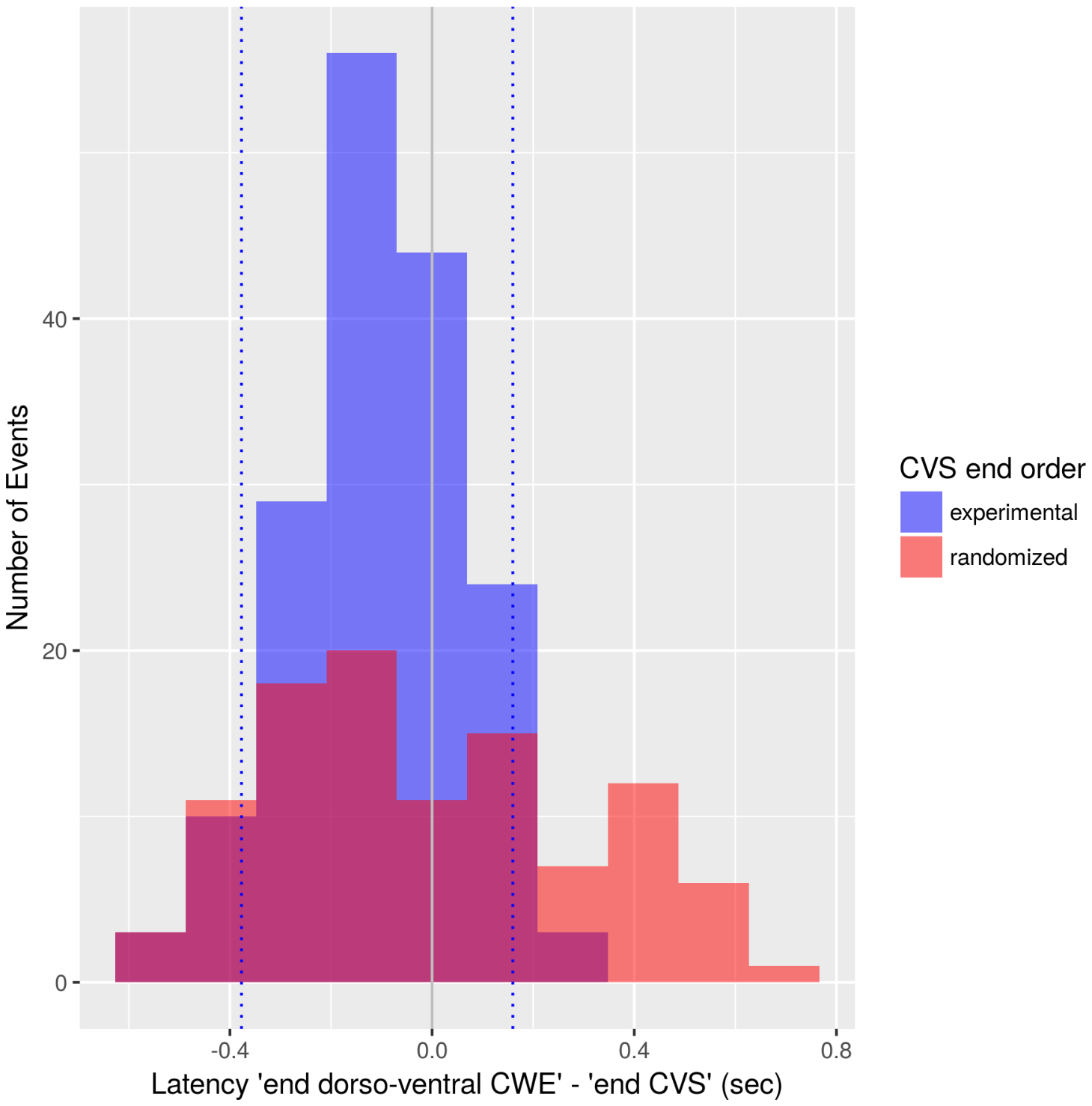}
\end{center}

\caption{
Histograms of latencies between the end of ventro-dorsal \glspl{CWE} and the
termination of \glspl{CVS}. Save format at in
Figure~\ref{fig:histLatenciesInitiationPosSpeedCWEs}.
90\% of \glspl{CWE} started between 0.34~sec before and 0.19~sec after the
termination of a \gls{CVS}.
The mean latency in the experimental dataset was -0.11~sec, which was
significantly smaller than chance (p\textless 1e-4, permutation test,
Section~\ref{sec:permutationTests}), suggesting that on
average \glspl{CWE} ended before the termination of productions of
\glspl{CVS}.
The variance of latencies in the experimental dataset was 0.03~sec$^2$, which
was significantly smaller chance (p\textless 1e-4, permutation test,
Section~\ref{sec:permutationTests}), suggesting that on average
\glspl{CWE} ended closer to the termination of productions of \glspl{CVS} than
expected by chance.
}

\label{fig:histLatenciesTerminationPosSpeedCWEs}
\end{figure}

Out of 92 ventro-dorsal \glspl{CWE}, 92 of them (100\%) overlapped
with the production of a \gls{CVS}.
For 90\% of these overlapping \glspl{CWE} their start time occurred between
-0.40 and 0.18~seconds from the start time of the overlapping \gls{CVS} (dotted
blue lines in Figure~\ref{fig:histLatenciesInitiationPosSpeedCWEs}),
and their end time occurred between -0.38 and 0.16 from the end time of the
overlapping \gls{CVS} (dotted blue lines in
Figure~\ref{fig:histLatenciesTerminationPosSpeedCWEs}).
That is, start and end times of ventro-dorsal \glspl{CWE} were aligned
to start and end times of productions of \glspl{CVS} with a sub-second
precision.

We tested the hypothesis that this alignment did not occurred by chance by
comparing statistics on start and end times in experimental datasets and in
surrogate datasets reflecting the null hypothesis of no alignment between
\glspl{CVS} and overlapped ventro-dorsal \glspl{CWE}
(Section~\ref{sec:surrogateDatasets}).
The mean of differences between start (end) times of ventro-dorsal \glspl{CWE}
minus start (end) times of overlapping \glspl{CVS} productions was -0.06~sec
(-0.11~sec), which was significantly smaller than chance, p=0.0185
(p\textless 1e-4) (permutation test, Section~\ref{sec:permutationTests}).
That is, on average \glspl{CWE} started and ended significantly before
overlapping \glspl{CVS}.
The variance of differences between start (end) times of ventro-dorsal
\glspl{CWE} minus start (end) times of \glspl{CVS} productions was
0.03~sec$^2$ (0.03~sec$^2$), which was significantly smaller than chance,
p\textless 1e-4 (p\textless 1e-4) (permutation test,
Section~\ref{sec:permutationTests}).
Thus, the alignment between start and end times of overlapping \glspl{CWE} and
\glspl{CVS} was significantly more precise than expected by chance.

\section{Discussion}

This is the third article in a series characterizing \glspl{TW} related to the
production of \glspl{CVS}. 
In \citet{rapelaInPrepTWsInSpeech} we showed for the first time that during
the production of \glspl{CVS} field potentials over speech processing brain
regions are spatio-temporally organized as \glspl{TW}.
In \citet{rapelaInPrepSyncTWs} we showed that these \gls{TW} are precisely
synchronized, in dynamical systems terms, to the initiation of the production
of \glspl{CVS}.
Here we demonstrated that these \glspl{TW} tend to appear before the start of
the production of \glspl{CVS} and they tend to disappear before the end of
these productions, with a sub-second precision.
To our knowledge, this is the first report of \glspl{TW} related to the
production of speech, and the first demonstration of a behavioral correlate of
mesoscale \glspl{TW} in behaving humans.

In future research we will investigate the relation between features of these
\glspl{TW} (e.g., velocity, propagation direction, duration) and phonetic
features of produced \glspl{CVS}. We will also study possible neural mechanisms
generating these reversing direction \glspl{TW} and their synchronization to
produced \glspl{CVS}.

\section{Methods}
\label{sec:methods}

We used the methods in \citet{pattenEtAl12} for the detection of traveling
waves and for the calculation of their speed. 

\subsection{Wave Event}
\label{sec:we}

\begin{figure}
\centering
\includegraphics[width=4in]{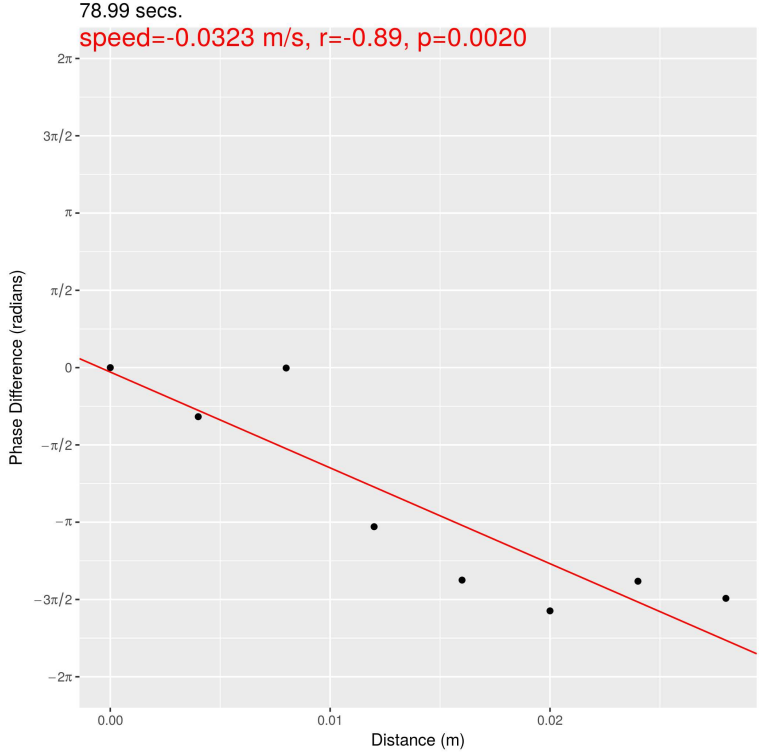}

\caption{
Detection of a Wave Event at time 78.99 sec between electrodes 135 and 142,
using electrode 142 at reference electrode.  The abscissa plots distances of
electrodes 142 to 135 to the reference electrode. The ordinate plots phase
difference between electrodes 142 to 135 and the reference electrode.  There
is a significant linear relation between phase differences and distances
(skipped Pearson correlation coefficient r=0.89; p=0.002, permutation
test) indicating a \gls{WE} at time 78.99~sec.
}

\label{fig:exampleWE}
\end{figure}

We studied \glspl{TW} between electrodes 135 and 142 in the ventral
sensorimotor cortex (Figure~\ref{fig:grid}). We used electrode 142 as
reference electrode.  We say that a wave event (\gls{WE}) occurred at time $t$
when there is a significant linear relation (p\textless0.01 and
$|r|$\textgreater0.85) between phase differences of electrodes with respect to
the reference electrode and distances of these electrodes to the reference
electrode.  Figure~\ref{fig:exampleWE} illustrates the detection of a \gls{WE}
at recording time 78.99 sec. 

In the chain of electrodes 142-135, higher numbered electrodes are more dorsal
than lower ones (e.g.; electrode 142 is dorsal to electrode 141, electrode 141
is dorsal to electrode 140, \ldots). A positive slope in  the best fitting
line in the plot Phase Difference vs Distance (red line in
Figure~\ref{fig:exampleWE}) indicates a \gls{TW} propagating in the
dorso-ventral direction from electrode 142 to electrode 135. Similarly, a
negative slope indicates a \gls{TW} propagating in the ventro-dorsal direction
from electrode 135 to electrode 142.

\subsection{Speed}
\label{sec:speed}

The slope of the best fitting line in the Phase Difference vs.\ Distance plot
gives the wave number, $k$, of the \gls{TW} and the speed, $v$, of the
\gls{TW} is $v=\frac{2*pi*f}{k}$, where $f$ is the \gls{TW} frequency. Blue
dots in Figure~\ref{fig:significantSpeeds} indicate \glspl{WE} occurring at
different times and with different speeds.

\subsection{Contiguous Wave Event}
\label{sec:cwe}

A Contiguous Wave Event~\gls{CWE} is a sequence of uninterrupted Wave Events,
such that \glspl{WE} were detected in all samples times from the sample time
of the first \gls{WE} to the sample time of the last \gls{WE}. The start/end
of a \glspl{CWE} in Figure~\ref{fig:significantSpeeds} are represented by
blue/red vertical lines, respectively.

\subsection{Surrogate Datasets}
\label{sec:surrogateDatasets}

Datasets used in the current analysis comprise voltage recordings from the 256
electrodes in the grid plus a transcription file indicating start and end
times of productions of \glspl{CVS}.
A surrogate dataset  contained identical voltage recordings as the original
dataset plus a modified transcription file, where the start and end times of
productions of \glspl{CVS} were randomized. Thus, in surrogate datasets there
should not be synchronization between \glspl{CVS} and \glspl{CWE}.

The modified transcription file randomized  start and end times of all
\glspl{CVS} in the original transcription file, but kept unchaged their
durations, as explained next.
The initiation time of the first \gls{CVS} in the modified transcription file
was that in the original transcription file plus a Gaussian random variable
with mean zero and standard deviation 0.5.
The termintation file of the first \gls{CVS} in the modified transcription
file was the ranomized initiation time plus the duration of the first
\gls{CVS} in the original transcription file.
The initation time of the second \gls{CVS} in the modified transcription file
was the termination time of the first \gls{CVS} in the modified transcription
file plus a random variable with mean and standard deviation equal to the mean
and standard deviation of inter-\gls{CVS}-intervals in the original
transcription file.
This procedure was repeated to generate randomized initiation and termination
times for all \glspl{CVS} in the original transcription file.

\subsection{Permutation Tests}
\label{sec:permutationTests}

We performed one-sided permutation tests (1) to asses the reliability of
proportions of \glspl{WE} ocurring in a given time interval (e.g.; the
proportion of ventro-dorsal \glspl{WE} ocurring between 0.3~sec before the
start of the production of a \gls{CVS} and the end of this production) and (2)
to test the significance of the alignment between the start/end time of a
\glspl{CWE} and the initiation/termination time of the production of a
\glspl{CVS}.
In both cases we first calculated a statistic on the original dataset (e.g.,
we computed the proportion of ventro-dorsal \glspl{WE} ocurring between
0.3~sec before the start of the production of a \gls{CVS} and the end of this
production in the original dataset). We then computed 2000 surrogate datasets
(as described in Section~\ref{sec:surrogateDatasets}), and we calculated the
previous statistic in each of the surrogate datasets. We thus generated a
collection of 2000 surrogate statistics. The p-value returned by a permutation
test was the proportion of surrogate statistics smaller (for left-sided tests)
or larger (for right-sided tests) than the test statistic on the original
dataset.

\section{Acknowledgments}

We thank Dr.\~Edward Chang for sharing the \gls{ECoG} recordings.

\printglossary

\bibliographystyle{plainnatNoNote}
\bibliography{travelingWaves,speech}

\begin{thebibliography}{7}
\providecommand{\natexlab}[1]{#1}
\providecommand{\url}[1]{\texttt{#1}}
\expandafter\ifx\csname urlstyle\endcsname\relax
  \providecommand{\doi}[1]{doi: #1}\else
  \providecommand{\doi}{doi: \begingroup \urlstyle{rm}\Url}\fi

\bibitem[Benucci et~al.(2007)Benucci, Frazor, and Carandini]{benucciEtAl07}
Andrea Benucci, Robert~A Frazor, and Matteo Carandini.
\newblock Standing waves and traveling waves distinguish two circuits in visual
  cortex.
\newblock \emph{Neuron}, 55\penalty0 (1):\penalty0 103--117, 2007.

\bibitem[Muller et~al.(2016)Muller, Piantoni, Koller, Cash, Halgren, and
  Sejnowski]{mullerEtAl16}
Lyle Muller, Giovanni Piantoni, Dominik Koller, Sydney~S Cash, Eric Halgren,
  and Terrence~J Sejnowski.
\newblock Rotating waves during human sleep spindles organize global patterns
  of activity that repeat precisely through the night.
\newblock \emph{eLife}, 5:\penalty0 e17267, 2016.

\bibitem[Patten et~al.(2012)Patten, Rennie, Robinson, and Gong]{pattenEtAl12}
Timothy~M Patten, Christopher~J Rennie, Peter~A Robinson, and Pulin Gong.
\newblock Human cortical traveling waves: dynamical properties and correlations
  with responses.
\newblock \emph{PLoS One}, 7\penalty0 (6):\penalty0 e38392, 2012.

\bibitem[Prechtl et~al.(1997)Prechtl, Cohen, Pesaran, Mitra, and
  Kleinfeld]{prechtlEtAl97}
JC~Prechtl, LB~Cohen, B~Pesaran, PP~Mitra, and D~Kleinfeld.
\newblock Visual stimuli induce waves of electrical activity in turtle cortex.
\newblock \emph{Proceedings of the National Academy of Sciences}, 94\penalty0
  (14):\penalty0 7621--7626, 1997.

\bibitem[Rapela(2016)]{rapelaInPrepTWsInSpeech}
Joaqu\'{i}n Rapela.
\newblock Entrainment of traveling waves to rhythmic motor acts, 2016.
\newblock URL \url{http://arxiv.org/abs/1606.02372}.

\bibitem[Rapela(2017)]{rapelaInPrepSyncTWs}
Joaqu\'{i}n Rapela.
\newblock Rhythmic production of consonant-vowel syllables synchronizes
  traveling waves in speech-processing brain regions, 2017.
\newblock URL \url{https://arxiv.org/abs/1705.01615}.

\bibitem[Rubino et~al.(2006)Rubino, Robbins, and Hatsopoulos]{rubinoEtAl06}
Doug Rubino, Kay~A Robbins, and Nicholas~G Hatsopoulos.
\newblock Propagating waves mediate information transfer in the motor cortex.
\newblock \emph{Nature Neuroscience}, 9\penalty0 (12):\penalty0 1549--1557,
  2006.

\end{thebibliography}

\end{document}